\documentclass{article}
\usepackage{cite}
\usepackage{graphicx}
\usepackage{dcolumn}

\begin{document}

\date{}
\author{Francisco M. Fern\'{a}ndez\thanks{%
fernande@quimica.unlp.edu.ar} \\
INIFTA, DQT, Sucursal 4, C.C 16, \\
1900 La Plata, Argentina}
\title{Comment on: ``The asymptotic iteration method revisited'' [J. Math. Phys.
61, 033501 (2020)]}
\maketitle

\begin{abstract}
In this comment we show that the eigenvalues of a quartic anharmonic
oscillator obtained recently by means of the asymptotic iteration method may
not be as accurate as the authors claim them to be.
\end{abstract}

In a recent paper Ismail and Saad\cite{IS20} revisited the asymptotic
iteration method (AIM) with the purpose of deriving conditions for its
validity. They first discussed some exactly-solvable textbook examples and
later applied the approach to the quartic anharmonic oscillator
\begin{equation}
\psi ^{\prime \prime }(x)+\left( x^{2}+Ax^{4}\right) \psi (x)=E\psi (x).
\label{eq:ANHO}
\end{equation}
They claimed to have obtained several eigenvalues ``accurate to fifty
decimals'' for $A=0.1$.

Our earlier experience with the AIM suggested that this approach
is less reliable and less accurate than other alternative
approaches\cite{F04,AF06,ACF06}, even with the improvement of an
adjustable parameter\cite{F04}. For this reason we were extremely
surprised by the accuracy attained by Ismail and Saad\cite{IS20}.
Since these authors did not show a convergence test in their paper
we decided to test their results by means of the Riccati-Pad\'{e}
method (RPM) that provides tight upper and lower bounds in the
case of the quartic anharmonic oscillator\cite{FMT89}.

Before carrying out the numerical calculation we outline some
theoretical results\cite{F04,AF06,ACF06} that appear to have been
overlooked by Ismail and Saad\cite {IS20}. The AIM is commonly
applied to the differential equation of second order
\begin{equation}
y^{\prime \prime }(x)=\lambda _{0}(x)y^{\prime }(x)+s_{0}(x)y(x),
\label{eq:diff_eq}
\end{equation}
that can be factorized into
\begin{equation}
\left[ \frac{d}{dx}+a(x)\right] \left[ \frac{d}{dx}+b(x)\right] y(x)=0,
\label{eq:diff_eq_factor}
\end{equation}
provided that the functions $a(x)$ and $b(x)$ satisfy
\begin{equation}
a(x)+b(x)=-\lambda _{0}(x),\;b^{\prime }(x)+a(x)b(x)=-s_{0}(x).
\label{eq:eqs_a,b}
\end{equation}
It follows from these equations that $b(x)$ is a solution to the Riccati
equation
\begin{equation}
b^{\prime }(x)-b(x)^{2}-\lambda _{0}(x)b(x)+s_{0}(x)=0.  \label{eq:Riccati_b}
\end{equation}

If we define
\begin{equation}
z(x)=y^{\prime }(x)+b(x)y(x),  \label{eq:z(x)}
\end{equation}
then equation (\ref{eq:diff_eq_factor}) becomes
\begin{equation}
z^{\prime }(x)+a(x)z(x)=0.  \label{eq:diff_eq_z(x)}
\end{equation}
Therefore, we can solve equation\ (\ref{eq:diff_eq_z(x)}) for $z(x)$ and
insert the result into equation (\ref{eq:z(x)}) that is then solved for $y(x)
$. In this way we obtain
\begin{eqnarray}
y(x) &=&e^{-\int^{x}b(x^{\prime })dx^{\prime }}\left\{
C_{2}+C_{1}\int^{x}e^{-\int^{x^{\prime }}\left[ b(x^{\prime \prime
})-a(x^{\prime \prime })\right] dx^{\prime \prime }}dx^{\prime }\right\} ,
\nonumber \\
&=&e^{-\int^{x}b(x^{\prime })dx^{\prime }}\left\{
C_{2}+C_{1}\int^{x}e^{\int^{x^{\prime }}\left[ \lambda _{0}(x^{\prime \prime
})+2b(x^{\prime \prime })\right] dx^{\prime \prime }}dx^{\prime }\right\} .
\label{eq:y(x)_sol}
\end{eqnarray}
where we have used the first of the two equations
(\ref{eq:eqs_a,b}). Notice that equation (\ref{eq:y(x)_sol}) is a
general result completely independent of any particular approach
like the AIM. Besides, the Riccati equation (\ref {eq:Riccati_b})
is commonly omitted in most applications of the AIM (see the paper
by Ismail and Saad\cite{IS20} and references therein).

Ismail and Saad\cite{IS20} showed some interest in differential equations of
second order with constant coefficients. When both $\lambda _{0}(x)$ and $%
s_{0}(x)$ are constant, then $b(x)$ is also a constant and the
Riccati equation (\ref{eq:Riccati_b}) reduces to the quadratic
equation $b^{2}+\lambda _{0}b-s_{0}=0$ that provides two roots.
Upon inserting any of these roots into equation
(\ref{eq:y(x)_sol}) we obtain the exact solution to the
differential equation. It is worth noticing that equation
(\ref{eq:y(x)_sol}) is valid for distinct ($s_{0}\neq -\lambda
_{0}^{2}/4$) or equal ($s_{0}=-\lambda _{0}^{2}/4$) roots.

Let us now focus on the calculation of the eigenvalues of the
quartic anharmonic oscillator (\ref{eq:ANHO}). In order to make
this comment sufficiently self-contained we outline the main ideas
of the RPM. In the case of the Schr\"{o}dinger equation with an
even potential $V(x)$ we define $\Phi (x)=x^{-s}\psi (x)$, where
$s=0$ or $s=1$ for even or odd states, respectively. Then, we
expand the logarithmic derivative $f(x)=-\Phi ^{\prime }(x)/\Phi
(x)$ in a Taylor series about $x=0$ ,
$f(x)=f_{0}x+f_{1}x^{3}+\ldots $ and obtain the Hankel
determinants $H_{D}^{d}(E)$ with matrix elements $f_{i+j+d-1}$,
$i,j=1,2,\ldots ,D$. It was proved that there are sequences of
roots $E^{[D,d]}$, $D=2,3,\ldots $ of
$H_{D}^{d}(E)=0$ that converge towards the actual eigenvalues from below ($%
d=0$) and above ($d=1$)\cite{FMT89}. In this way one obtains increasingly
accurate lower and upper bounds, respectively.

Table~\ref{tab:UBLB} shows the remarkable (in fact it is
exponential) rate of convergence of the bounds for the ground
state of the quartic anharmonic oscillator (\ref{eq:ANHO}) with
$A=0.1$. Table~\ref{tab:compara} compares present bounds with the
results of Ismail and Saad\cite{IS20}. Our bounds suggest that
more than half of the significant figures reported by those
authors may not be correct.

Curiously, the thirteen significant digits reported by Ismail and Saad\cite
{IS20} for the case $A=2$ are consistent with our more accurate bounds $%
E^{[15,0]}=1.60754130246854753870817192941<E^{[15,1]}=1.60754130246854753870817192948
$.

Summarizing: more than half of the decimal figures shown by Ismail and Saad%
\cite{IS20} for the quartic anharmonic oscillator with $A=0.1$ do not appear
to be correct. It may be due to lack of convergence of the AIM or to round
off errors caused by insufficient digits in the calculation.

\begin{table}[tbp]
\caption{Convergence of the Lower bounds ($d=0$) and upper bounds ($d=1$)
for the ground state of the quartic anharmonic oscillator with $A=0.1$ }
\label{tab:UBLB}
\begin{center}
\par
\begin{tabular}{ccc}
\hline
$D$ & $d=0$ & $d=1$ \\ \hline
2 & 1.065165589106464508643143086785809633638 &
1.065291556141124441135238488833718516162 \\
3 & 1.065285181369961298428253752818854854099 &
1.065285528386575099263974255031454998383 \\
4 & 1.065285508412319469577830463652960342502 &
1.065285509614182897898433926644472417431 \\
5 & 1.065285509539192592585488356076943661193 &
1.065285509544015954376247941103453239435 \\
6 & 1.065285509543697581134961945367065150205 &
1.065285509543719071409991810845081902423 \\
7 & 1.065285509543717592126285592017093474034 &
1.065285509543717695730809861353737125740 \\
8 & 1.065285509543717688361781653356859779602 &
1.065285509543717688893236061269397044334 \\
9 & 1.065285509543717688854423603909817897904 &
1.065285509543717688857290648469407292386 \\
10 & 1.065285509543717688857076639211237627062 &
1.065285509543717688857092767847161100393 \\
11 & 1.065285509543717688857091541516086073081 &
1.065285509543717688857091635527212915542 \\
12 & 1.065285509543717688857091628265134465940 &
1.065285509543717688857091628830109065701 \\
13 & 1.065285509543717688857091628785862167126 &
1.065285509543717688857091628789349091100 \\
14 & 1.065285509543717688857091628789072686577 &
1.065285509543717688857091628789094718040 \\
15 & 1.065285509543717688857091628789092952804 &
1.065285509543717688857091628789093094939
\end{tabular}
\end{center}
\end{table}

\begin{table}[tbp]
\caption{Lower bound, result of Ref.\protect\cite{IS20} and upper bound for
the first states of the quartic anharmonic oscillator with $A=0.1$}
\label{tab:compara}
\begin{center}
\par
\begin{tabular}{l}
\multicolumn{1}{c}{$n=0$} \\
1.065285509543717688857091628789092952804 \\
1.06528550954371768885687796202255128719116328284144 \\
1.065285509543717688857091628789093094939 \\
\multicolumn{1}{c}{$n=1$} \\
3.306872013152913507128121684692867756592 \\
3.30687201315291350712686699320208560948231024667621 \\
3.306872013152913507128121684692869154624 \\
\multicolumn{1}{c}{$n=2$} \\
5.747959268833563304733503118475917140926 \\
5.74795926883356330473447484696869480558234499767423 \\
5.747959268833563304733503118477229464674 \\
\multicolumn{1}{c}{$n=3$} \\
8.352677825785754712155257734637775310436 \\
8.35267782578575471215441908268140025484171928837895 \\
8.352677825785754712155257734644178775630
\end{tabular}
\end{center}
\end{table}

\end{document}